\documentclass[12pt,preprint]{aastex}
\usepackage{psfig}
\def\HI {H\kern0.1em{\sc i}} 
\def\water {H$_{\rm 2}$O}
\def\txs {TXS\,2226{\tt -}184}

\def\deg{$^{\circ}$}
\def\kms{km s$^{-1}$}

\begin{document}
\title{~~\\ ~~\\ Exploring the Nucleus of the Gigamaser Galaxy TXS\,2226{\tt -}184}
\shorttitle{The nucleus in \txs}
\shortauthors{Taylor et al.}
\author{G. B. Taylor\altaffilmark{1}, A. B. Peck\altaffilmark{2}, 
J. S. Ulvestad\altaffilmark{1}, \& C. P. O'Dea\altaffilmark{3} }

%\email{gtaylor@nrao.edu}
\altaffiltext{1}{National Radio Astronomy Observatory, P. O. Box 0, Socorro, NM
  87801, USA}
\altaffiltext{2}{Harvard Smithsonian Center for 
Astrophysics, SAO/SMA Project, P.O. Box 824, Hilo, HI 96721, USA}
\altaffiltext{3}{Space Telescope Science Institute, Baltimore, MD 21218, USA}

%\received{2000 March 24}
%\accepted{2000 May 9}
%\journalid{337}{15 January 1989}
%\articleid{11}{14}

\slugcomment{As accepted to the Astrophysical Journal}

\begin{abstract}

We present Very Long Baseline Array (VLBA) observations of the
Gigamaser galaxy TXS2226{\tt -}184 at 1.3 and 5 GHz.  These
observations reveal the parsec-scale radio structure of this Seyfert
galaxy with exceptionally luminous water maser emission.  The source
is found to be extended on scales of 10-100 pc with some embedded
compact sources, but has no readily identifiable flat-spectrum active
nucleus.  This morphology resembles that of the nearby compact
starburst galaxy Mrk~273, although no significant FIR emission has
been detected to support the starburst scenario.  The narrow (125
\kms) \HI\ absorption in TXS2226{\tt -}184 discovered with the VLA is
also detected with the VLBA.  This \HI\ absorption is distributed
across the extended emission, probably co-spatial with the water
masers.  The broad (420 \kms) line seen by the VLA is not detected,
suggesting that it arises from more extended gas which is absorbing
the emission beyond the central tens of parsecs.

\end{abstract}


\section{Introduction}

Galaxies with luminous water masers in their nuclear regions have been
keenly sought after in recent years (e.g., Braatz, Wilson \& Henkel
1997) following the discovery of the masers in close orbit around the
nucleus of NGC~4258 (Miyoshi et al. 1995, Greenhill et al. 1995).
This system provided some of the earliest and strongest evidence for
the presence of a supermassive black hole in the nucleus and led to a
direct distance measurement to NGC~4258 which has helped to refine the
extragalactic distance scale (Herrnstein et al. 1999).

The water maser emission in \txs\ was discovered by Koekemoer et al.\
(1995) using the Effelsberg telescope.  This system hosts the most
luminous known extragalactic H$_2$O maser source, a so-called
``gigamaser'', with an isotropic luminosity in the 22 GHz line of 6100
L$_\odot$ (Koekemoer et al.\ 1995). The water maser emission from
\txs\ is fairly broad, with a FWHM of 88 km s$^{-1}$, in contrast to
most known extragalactic water masers with linewidths of only a few km
s$^{-1}$.  VLBI observations by Ball et al. (2004, in preparation)
reveal that the masers are distributed in clumps that trace a disk
oriented in position angle $-$65\deg, about 30 degrees tilt away from
the major axis of the galaxy.  One blueshifted maser appears
significantly offset from this disk, which may indicate that it is
associated with the jet.

Recent HST observations by Falcke et al.\ (2000) classify
the galaxy as a highly inclined spiral and reveal a dust lane cutting
across the nucleus.  Falcke et al.\ (2000) also present VLA
observations at 8.4 GHz showing that the radio emission is compact
($<$ 1\arcsec), symmetric, and has an axis perpendicular to the dust
lane.  No larger-scale diffuse emission is present to the sensitivity
limits of the NRAO VLA Sky Survey (NVSS) at about 2 mJy beam$^{-1}$ (Fig.~2;
Condon et al.\ 1998).

There is a fair amount of overlap in the occurance ($\sim$40\%)
between extragalactic H$_2$O maser systems and \HI\ found in
absorption (Taylor et al. 2002).  The \HI\ can be used to study the
dynamics near the central engine of active galaxies, illuminating the
process of accretion and jet propagation near a massive black hole.
For this reason Taylor et al. (2002) undertook a study of the \HI\ gas
in \txs\ with the VLA and discovered that it consists of two
components -- one with a width of 125 km s$^{-1}$, and another broader
feature of width 420 km s$^{-1}$.  Both velocity components are found
toward the compact radio source in the nucleus of the galaxy,
co-spatial within the uncertainties with the water masers.  Taylor et
al. suggested that the narrow line might be indicative of an
interaction between the radio jet and the surrounding material.

In this study we make use of VLBI observations that resolve the radio
continuum in the central 0.5 kpc, and also spatially resolve the \HI\
absorption.  We use this information to
better characterize the nature of this system and to determine the
powering mechanism for the exceptionally luminous maser emission.

Throughout this discussion, we assume H$_{0}$=71 km s$^{-1}$
Mpc$^{-1}$, $\Omega_M$ = 0.27, and $\Omega_{\Lambda}$= 0.73, resulting in
a linear to angular scale ratio of 0.496 kpc arcsecond$^{-1}$
\footnote{Derived using E.L. Wright's cosmology calculator at
http://www.astro.ucla.edu/~wright/CosmoCalc.html.}.

% luminosity distance is 107.6 Mpc
  
\section{The VLBA Observations}

The observations were made with the National Radio Astronomy 
Observatory\footnote{The National Radio Astronomy Observatory is operated
by Associated Universities, Inc., under cooperative agreement with the
National Science Foundation.} Very Long Baseline Array (VLBA) and Robert C. Byrd Green Bank
Telescope (GBT) at a center frequency of 1386 MHz on 2002 December 14
and 15.  A total of 8.4 hours were obtained on source using 256
channels across a 16 MHz band to provide a resolution of 14 \kms.
Both right and left circular polarizations were observed.  Phase
calibration was obtained by short (1 min) observations of the nearby
(1.98 degrees distant), moderately strong (0.25 Jy) calibrator
J2236{\tt -}1706 every 3 minutes.  Bandpass calibration was provided
by observations of J2253+1608.

Observations at 4982 MHz were carried out with the VLBA alone on
2002 December 16.  A total of 2.2 hours were obtained on source 
with a 32 MHz bandwidth observing 
in right circular polarization only.  Phase referencing was once
again performed by switching to J2236{\tt -}1706 every 3 minutes for
1 minute.  We also checked the atmospheric coherence by observing
J2236$-$1433 once every 38 minutes.

\section{Results}

\subsection{MERLIN Continuum Images}

Data at 5 GHz from observations of the \txs\ field were obtained from
the Multi-Element Radio-Linked Interferometer Network (MERLIN)
archive.  These data consisted of two observing sessions, one on 1999
February 3, and another on 1999 March 3.  In each session 6 antennas
participated for a total of 4.2 and 4.9 hours on source in February
and March respectively.  Calibration of these data was performed in the
standard fashion in AIPS using the nearby calibrator J2232{\tt -}1659
observed every 7 minutes.  
Data from the two days were
combined and then imaged using {\sc Difmap}.  The rms noise in the
total intensity image is 0.085 mJy beam$^{-1}$.

At the MERLIN resolution of 40 $\times$ 179 mas \txs\ is dominated by
a compact core, but shows jet-like extensions to the northwest and
southeast (Fig.~1).  The compact core has a peak flux density at 5 GHz
of 13.6 mJy.  The orientation of the jets at 144\deg\ is in excellent
agreement with the the elongation angle of 145\deg\ seen in the 22 GHz
VLA observations (Taylor et al. 2002) with resolution 390 $\times$ 210
mas.  In the MERLIN image we find a total of 31.1 $\pm$ 0.93 mJy
compared to 32.2 $\pm$ 0.98 detected by the VLA (Taylor et al.\ 2002).
A little less than half of this emission, 13.6 mJy, is in the compact
core with a size of $<$45 mas.

\subsection{VLBA Continuum Images}

   Starting with the phase referenced VLBA+GBT image at 1.4 GHz 
from the standard
calibration in AIPS, we performed phase self-calibration in {\sc
Difmap}.  In Fig.~2 we show the 1.4 GHz continuum emission from \txs\
at resolutions of 20 $\times$ 12 mas and 45 mas.  Both images were
made using natural weighting and applying a taper to downweight the
longest spacings.  Only 35 mJy, about half the flux density of the
73.3 mJy compact VLA core, is successfully recovered in these images.
Besides a bright and resolved region of diameter 0.1 arcseconds, there
is diffuse emission to the northwest along a position angle of
$-$36\deg. This agrees well with the VLA and MERLIN orientations on
somewhat larger scales (see Fig.~1). The milliarcsecond-scale emission is broken
up into clumps, most likely as a result of difficulties of the clean
algorithm in recovering the extended emission.  A multi-resolution
clean was attempted in AIPS, but did not recover more flux density or
produce a better image.  The MERLIN image at 5 GHz shows diffuse
emission on scales of $\sim$0.3 arcseconds that we do not resolve with
the VLA, but probably over-resolve with the VLBA.  The rms noise in the VLBA
continuum image is 50 microJy/beam in the high resolution image and
100 microJy/beam in the 45 mas image owing to the heavier taper
applied.

  The rms noise at 5 GHz is 100 microJy/beam, and the peak in the map
is 500 microJy.  While there is a suggestion of some flux density on
the shortest baselines, no sources are reliably detected.  From our
phase referencing check source, J2236$-$1433, we estimate the
coherence to be greater than 50\%.  We place an upper limit of 1
mJy beam$^{-1}$ as the strongest unresolved source that could go undetected
at 5 GHz.  An attempt to combine the VLBA and MERLIN data was made,
but owing to the lack of signal on the VLBA baselines, this was
unsuccessful.

\subsection{The \HI\ Absorption}

  In Fig.~3 we show \HI\ spectra for four of the brightest regions in
the high resolution image.  The continuum level has been subtracted
from the spectral line cube.  The spectra have been Hanning smoothed
and averaged to a velocity resolution of 23.98 \kms.  The absorption
peaks in all spectra at 7500 \kms\ in the heliocentric frame. In the
region of highest SNR the absorption has a depth of 2.6 mJy, and a
FWHM of 125 \kms.  There is a hint of an additional component around
7700 \kms\ in all but the weakest profile.

%south to north  in 45 mas beam
%A:  100 75 ~3.5 mJy continuum
%B:  116 91 13.0 mJy
%C:  139 129 3.0 mJy
%
%south to north in 20 x 12 beam
%a:  104 81 1.93 mJy
%b:  111 91 3.60 mJy
%c:  120 91 2.82 mJy
%d:  141 128 2.40 

   We have created an image of the integrated optical depth over
the line by averaging 24 channels centered on the peak absorption
using the cube with 45 mas spatial resolution (Fig.~4).
We find a marginally significant increase in the optical depth from 
0.15 $\pm$ 0.04 to 0.34 $\pm$ 0.08.  

   Finally, we have searched for gradients in velocity by
averaging in the direction perpendicular to the major axis of
the source.  To do this we 
rotated the cube spatially by $-$55\deg\ and then averaged
over the source in declination.  A position-velocity plot from
this averaged cube is shown in Fig.~5.  No significant gradient
in velocity is seen.

% sqash in declination from pixel 97 to 146

\subsection{Location of the Neutral Hydrogen Gas}

Based on the VLA observations, Taylor et al. concluded that the
$\sim$420 \kms\ wide absorption feature with a depth of 5.6 mJy probably
results from neutral material associated with the atomic and molecular
torus thought to feed the active nucleus, and that the deeper
$\sim$125 \kms\ wide line in \txs\ could be indicative of an interaction
between the radio jet and surrounding material.  If the supposition
about the nature of the broad line were correct, then a compact
nucleus with at least 5.6 mJy should have been visible in the 1.4 GHz
continuum image.  The peak in the full resolution image (not shown) is
only 2.4 mJy.  From the absence of any broad-line absorption in our
VLBI spectra we conclude that the broad line originates from emission
on spatial scales of $\sim$0.3 arsec that are not probed by our VLBA
observations.  Similarly, a 1000 \kms\ wide blue-shifted line seen in the 
WSRT spectrum towards 3C\,293 (Morganti et al. 2003) was not detected
on the parsec-scale by Beswick et al. (2004).  Another example of 
broad lines resolving out on the parsec-scale can be found in 4C12.50 
(Morganti et al. 2004).  Morganti et al. (2003) speculate that 
the broad line arises from outflow of gas on kiloparsec-scales, and 
the same situation could be responsible for the broad line in \txs.  

The narrow (125 \kms) and deeper (12.3 mJy for the VLA)
\HI\ line in \txs\, has a similar width and velocity dispersion to the
water masers and could originate from the same region.  
The lack of any significant velocity 
gradient in the \HI\ observations is consistent with this 
interpretation. 

Another possibility is that we are seeing multiple \HI\ clouds from
throughout the spiral host galaxy along the line-of-sight toward the
nucleus.  In this case the \HI\ gas would not be associated with the
nucleus or the water masers.  The marginally significant change in
\HI\ opacity across the source (see Fig.~4) argues against this
possibility since the average \HI\ absorption from the host galaxy is
unlikely to change on scales as small as 100 pc.

\section{Nature of the Continuum Emission and Water Masers }

Observationally, extragalactic water masers can be separated into
three categories based on their line widths, luminosities and
distribution of masing regions.  One class, with narrow linewidths of
a few \kms, and isotropic luminosities $\sim$100 L$_\odot$, is most
frequently associated with the accretion disk around an AGN (e.g.,
NGC~4258 , Miyoshi et al.  1995).  The second, possibly less common,
class of masers is characterized by broader linewidths of $\sim$100
\kms, and luminosities in the range 100-3000 L$_\odot$ (e.g., Mrk 348,
Peck et al. 2003; NGC~1052, Claussen et al. 1998).  The third, recently identified class are weaker, $<$10
L$_\odot$, masers with narrow linewidths (e.g., NGC 2146, Tarchi et
al. 2002).  The criteria for characterizing the different classes of
masers are outlined in Peck et al (2004).

Assuming that the masing region completely covers the radio emitting
core at 22 GHz, Koekemoer et al. (1995) found that the required
maser amplification in \txs\ is a factor of $\sim$20.  If the maser
covers less of the continuum source then the amplification must
be even greater.  
In this section we consider three scenarios to explain the nature 
of the radio emission and the extragalactic water masers in \txs.

\subsection{Case 1: A Compact Starburst}

The clumpy, steep-spectrum radio morphology of \txs\ on the parsec
scale (Fig.~2) is reminiscent of the starburst galaxies Mrk~231 and
Mrk~273 (Carilli, Wrobel, \& Ulvestad 1998, Carilli \& Taylor 2000),
which also contain active galactic nuclei.  Given the steep spectrum
of the overall emission in \txs , the individual components could be
radio supernovae, or clusters of SNe. The H$_2$O emission and \HI\
absorption likewise are found to occur on similar spatial scales in
these prominent starburst systems.  

A significant problem with the starburst scenario is the low infrared
luminosity.  As listed in the NASA/IPAC Extragalactic Database (NED),
\txs\ is not detected by IRAS at 12, 25 and 100 microns, and is only
309 $\pm$ 59 mJy at 60 microns.  Given the distance to the source this
corresponds to a power at 60 microns of 4.15 $\times$ 10$^{23}$ W/Hz.
Scaling the starburst emission models of Rice et al. (1988) we get
$L_{IR} = 7.5 \times 10^9 L_{\odot}$.  This corresponds to a star
formation rate of just 0.178 M$_{\odot}$/yr (Heckman et al. 1990), far
below what is found in even the weakest starburst galaxies.  One can
also ask the question of how well this system obeys the well
established radio-far IR correlation (Condon 1992).  The integrated radio
power at 1.4 GHz is $P_{1.4} = 1.01 \times 10^{23}$ W/Hz.  We find
a logarithmic ratio of FIR to radio power,
Q=$-$2.3, compared to the average value of $+$2.3 $\pm$ 0.2 (Condon
1992).  At present, given the anomalous nature of these findings, we
cannot rule out the possibility that there is some error in the IRAS
measurement, so we plan to undertake new IR observations to confirm
the value.

That we could be seeing individual supernovae in \txs\ is unlikely
since Cas~A would have a flux density of only $\sim$1 $\mu$Jy at the
distance of \txs, although young radio supernovae might have powers
100 -- 1000 times that of Cas A (Weiler et al. 2002; Neff et
al. 2004).  However, we would expect them to have some reasonable flux
at 5 GHz if they are young enough to have such high brightness
temperature, so the lack of any detection at 5 GHz argues against the
supernova hypothesis.  The distribution of the radio emission
perpendicular to the major axis is also not readily explained by SNe.

It is possible that this source could be a ``post-starburst'' object in
which the diffuse radio emission was created by a starburst.
Koekemoer et al. (1995) refer to unpublished observations that suggest
an optical continuum dominated by a ``post-starburst'' stellar
population.  Assuming minimum energy conditions apply, the fairly high
field derived, $B_{me}$ = 630 $\mu$G, together with the observed
spectral index of $-$0.66 between 1.4 and 5 GHz, implies an age of the
radio source of only 14,000 years (Myers \& Spangler 1985).  If the
radio source is considerably older then some reacceleration process
must be at work.  Over the $\sim$10$^8$ years since the starburst one might
expect the radio source to have become much less collimated than we
currently observe.  We
conclude that the small amount of star formation taking place in \txs\
is unlikely to produce the majority of the radio continuum emission.
It is also difficult to see how this low amount of star formation
could provide the shocks necessary to create the conditions for
masing.

\subsection{Case 2: Amplification of a Background AGN}

The parsec-scale radio emission is elongated in the same direction as
the kiloparsec-scale emission, and is perpendicular to the major axis
of the host galaxy and the inner dust disk (Falcke et al. 2000).  The
symmetric structure and tight collimation of the outflow in \txs\
suggests that the radio continuum is produced by jets oriented at a
large angle to the line-of-sight propagating perpendicular to the
accretion disk.  Falcke et al. also find that the radio emission is
aligned with H$\alpha$ + [N II] emission and suggest that this Narrow
Line Region (NLR) is produced by the interaction between the radio jet
and the interstellar medium (Falcke, Wilson \& Simpson 1998).
Koekemoer et al (1995) note that the optical spectrum of the nucleus
has line ratios typical of a low-ionization nuclear emission region
(LINER) spectrum.  The optical spectrum and the presence of jets
suggest the presence of an active nucleus.

   An examination of the radio continuum emission reveals no obvious
compact, flat spectrum source readily identifiable with the nucleus.
At 5 GHz we can use the limiting flux density of 1 mJy to put an upper
limit on the core radio power at 5 GHz of $<$ 1.4 $\times$ 10$^{21}$
W/Hz.  This core power is lower than all but a few low luminosity radio
sources in the Complete Bologna Sky Survey (Giovannini et al. 2004)
despite the fact that the radio power of $P_{0.365} = 2.72 \times\
10^{23}$ W/Hz at 365 MHz is only about an order of magnitude below the
average total power in a complete sample of radio galaxies.  From this
we can conclude that the core is extremely underluminous.  This could
imply subrelativistic ejection and/or an angle very close to the plane
of the sky.  Alternatively, the core could be heavily obscured by
free-free absorption.  Since the free-free absorption from an external
screen decreases exponentially with frequency, further sensitive,
high-frequency VLBI observations might detect an obscured,
flat-spectrum nucleus.  Such a heavily absorbed nucleus might also
make a significant contribution to the total flux density at high
frequencies.  There is no indication of any flattening in the spectrum
up to 22 GHz (Taylor et al. 2002), from which we place a limit on the
contribution of any flat spectrum component of $<$10\% or $<0.7$ mJy.
Ball et al. (2004, in preparation) place a limit of $<$ 2 mJy on
any compact continuum emission at 22 GHz.
If the masers are limited to the central few parsecs around a
low-power core then the amplification needed will be $>$600.  Alternatively,
the X-ray illumination provided by the AGN may create a disassociation 
region and power the maser directly (Neufeld, Maloney, \& Conger 1994).  
Koekemoer et al. (1995) find that this model can account for the
maser luminosity of \txs\ if the illuminated disk is 2.5 -- 7.5 pc 
in radius.  Ball et al. (2004, in preparation) find a disk of radius 5 pc,
but do not rule out the possibility that some of the components
are associated with the radio jet.

\subsection{Case 3: Shocks Driven by a Jet}

   As already established in the previous section, it is likely that
there are bidirectional jets in \txs.  The peak radio flux density at
1.4 GHz in \txs\ is 3.6 mJy.  The brightness temperature of this
emission is 1.4 $\times 10^7$ K.  This, and the steep spectrum of the
integrated emission ($\alpha = -0.66$ from Taylor et al. 2002, where
$S_\nu \propto \nu^\alpha$) clearly indicates non-thermal emission.

   A possible explanation for the luminous water masers in \txs\ is
from shocks produced by a jet driving into a molecular cloud, though
if this is to explain all the masers then the jet needs to bend
through 60 degrees in order to line up with the jet seen on scales
of 10-1000 pc (Ball et al. 2004, in preparation).  The
situation may be comparable to that in Mrk~348, a Seyfert galaxy at a
distance of 62.5 Mpc which hosts an \water\ megamaser (Peck et
al. 2003).  In Mrk~348, the maser appears to be caused by the radio
jet impacting a molecular cloud within the central few parsecs of the
galaxy.  This interaction gives rise to an expanding bow-shock driven
into the cloud which has a velocity between 135 \kms\ and 0.5c in the
direction of jet propagation, and between 35 \kms\ and 300 \kms\ at
various points along the oblique edges (Peck et al. 2003).  This shock
generates a region of very high temperature, ($\le$10$^5$ K), which
dissociates the molecular gas and to some extent shatters the dust
grains expected to be present and/or evaporates their icy mantles.
Immediately following this shock, H$_2$ begins forming on the
surviving dust grains when the temperature has dropped to $\sim$1000 K,
and this in turn provides sufficient heating to stabilize the
temperature at $\sim$400 K, with gas densities of $\sim$10$^8$ cm$^{-3}$
(e.g., Mauersberger,
Henkel \& Wilson 1987; Elitzur 1995).  In Mrk~348, the pc-scale radio
jet is significantly brighter than that in \txs, but it is possible
that a weaker jet could result in stronger amplification than a powerful
jet, given that it is the regions of oblique shocks, with velocities
lower than 300 \kms, which provide post-shock conditions leading to a
significant volume of masing gas. 

This possibility has also been explored for
another Seyfert galaxy, NGC~1068.  A similar luminosity, linear radio
emission extending over $\sim$1'' is seen in NGC~1068 (Gallimore,
Baum, \& O'Dea 1996).  In this well-studied source Gallimore et
al. find a bending jet and local flattening of the spectral index
which lends support to this model.  

One of the chief diagnostics of this type of maser is the shape and
width of the emission line (Peck et al 2004), given that jet and
accretion disk geometry can be very difficult to decipher at large
distances using the distribution of \water\ spots and \HI\ absorption,
even with extremely high angular resolution.  The FWHM of the maser
emission in \txs\ is 88 \kms, much closer to the 130 \kms\ FWHM
measured in Mrk~348 than to the narrower lines associated with masers
known to arise in accretion disks.  Nonetheless, we look forward to
higher dynamic range VLBA and Expanded VLA images to assist in testing
this model in \txs.

\section{Conclusions}

Our observations show that on parsec scales the continuum emission
from an active nucleus in \txs\ is extremely weak.   Most of the 
emission originates from a jet which extends
over $\sim$100 pc.  \HI\ absorption is detected against this jet with a
marginal change in opacity, but no sign of rotation or outflow.  The
water masers may amplify either a weak nucleus, or extended jet
emission.  In either case, high amplification factors ($>600$) are
needed. 

Given the extreme luminosity of the water masers in \txs\ and their
broad line widths, it is tempting to ascribe this and similar systems
(e.g., Mrk 348, NGC 1052) with jets that drive shocks into molecular
clouds.  Our high-resolution radio continuum observations are
consistent with this picture.  Identifications of more systems, and
detailed studies of time-variability in the masers and continuum
emission are needed.

VLBI observations of the water masers will help to
understanding the nature of this system, though the preliminary
analysis indicates that the situation is complicated with  
both disk and jet masers present (Ball et al. 2004, 
in preparation).  Single dish observations
could also yield important clues by looking for rapid variations in
the flux density of the maser line, and correlating them with changes
in the continuum flux density to test the idea that the masers amplify
the background continuum.

\acknowledgments
We thank Y. Pihlstroem and an anonymous referee for insightful
suggestions.
This research has made use of the NASA/IPAC Extragalactic Database (NED)
which is operated by the Jet Propulsion Laboratory, Caltech, under
contract with NASA. This research has also made use of NASA's Astrophysics
Data System Abstract Service.

\clearpage

\bigskip
\clearpage

%Figure 1 -- Merlin Image 
\begin{figure}
\plotone{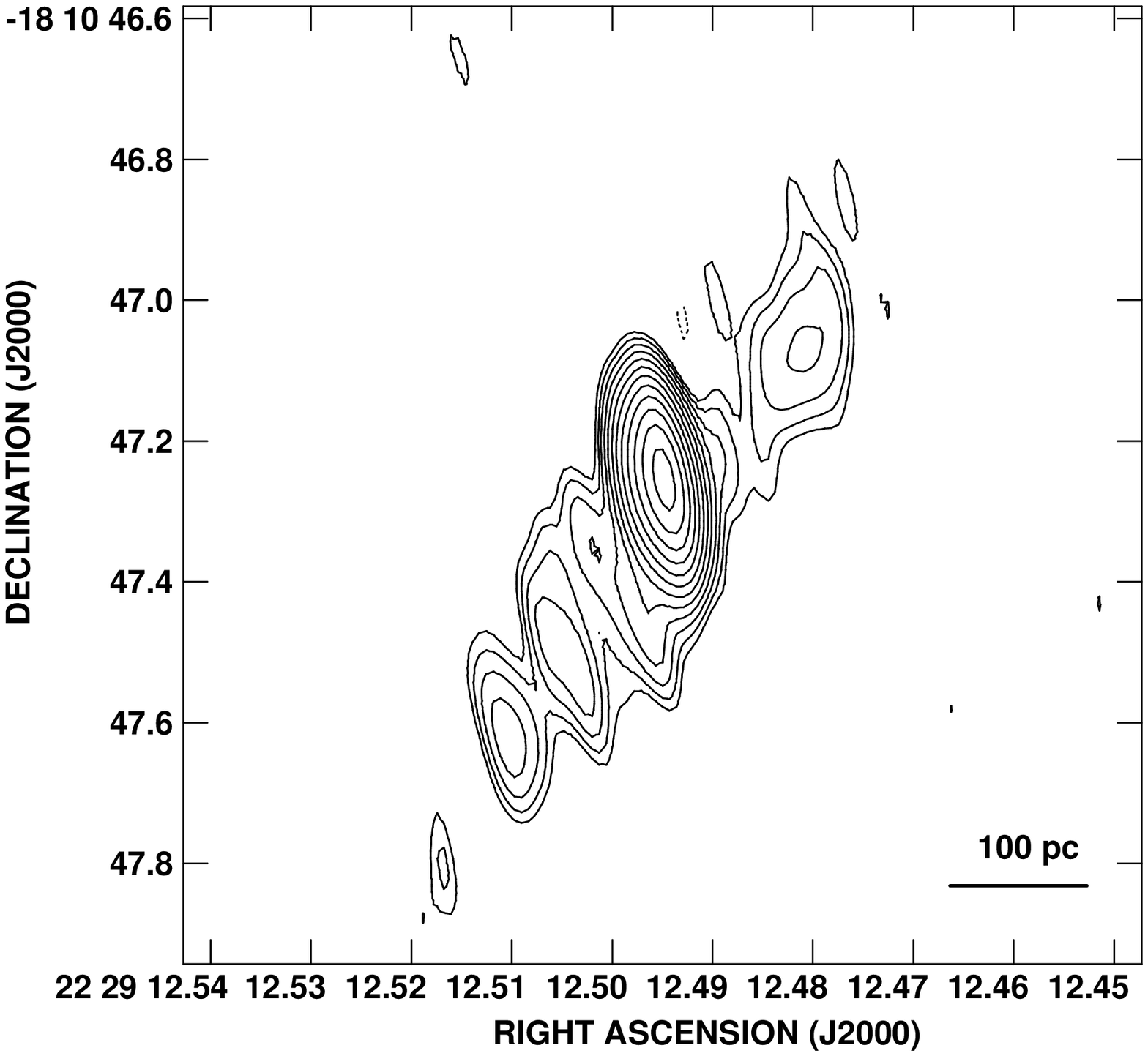}
\caption{The MERLIN image at 5.0 GHz.  The resolution is 179 $\times$ 41 mas
in position angle 12\deg.  Contours
are drawn at $-$0.26, 0.37, 0.51, ..., 12.8 mJy beam$^{-1}$
by $\sqrt{2}$ intervals with 
negative contours shown dashed.  The peak in the image is 13.6 mJy beam$^{-1}$.\label{fig1}}
\end{figure}

%Figure 2 -- Continuum images
\begin{figure}
\plottwo{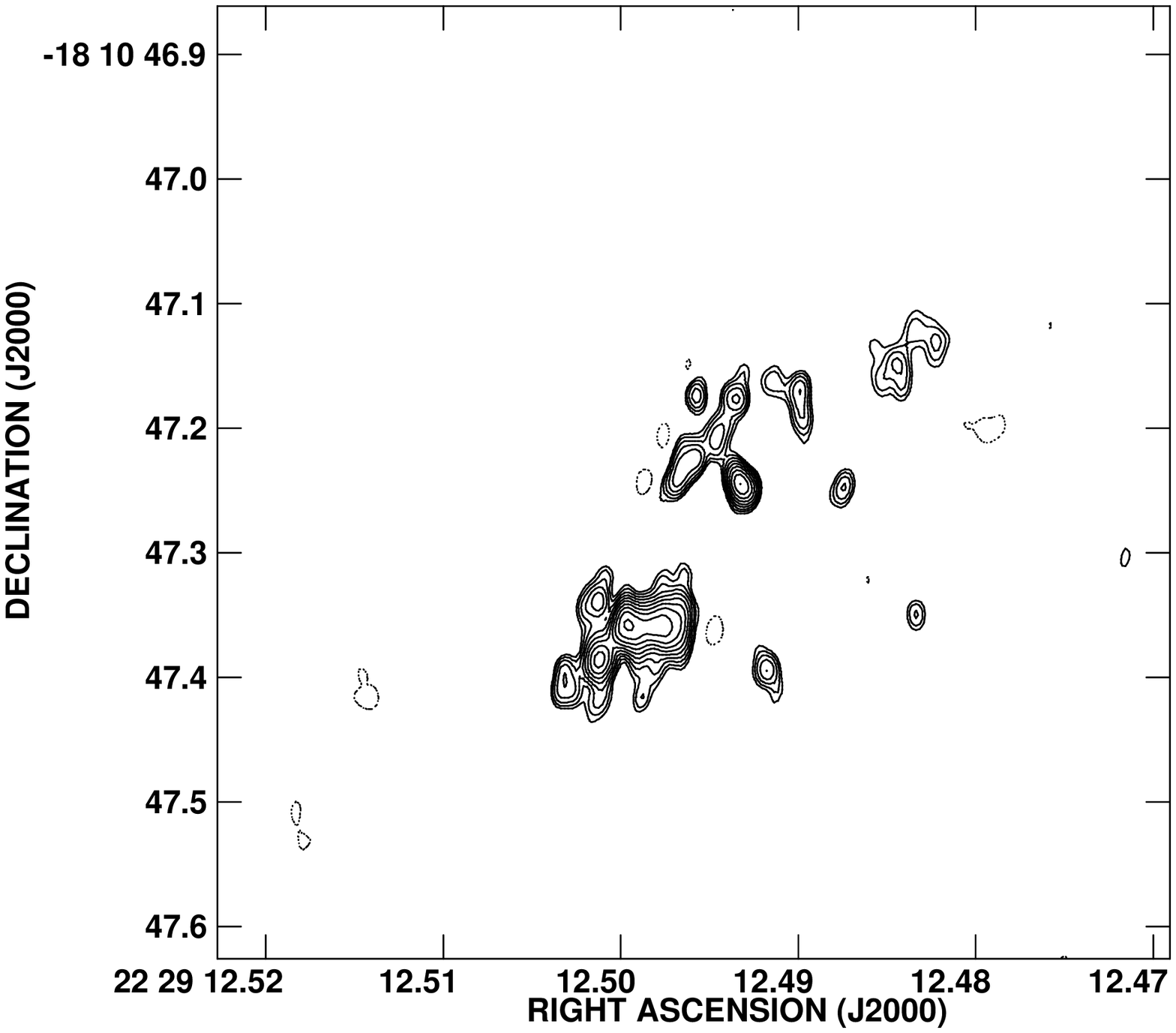}{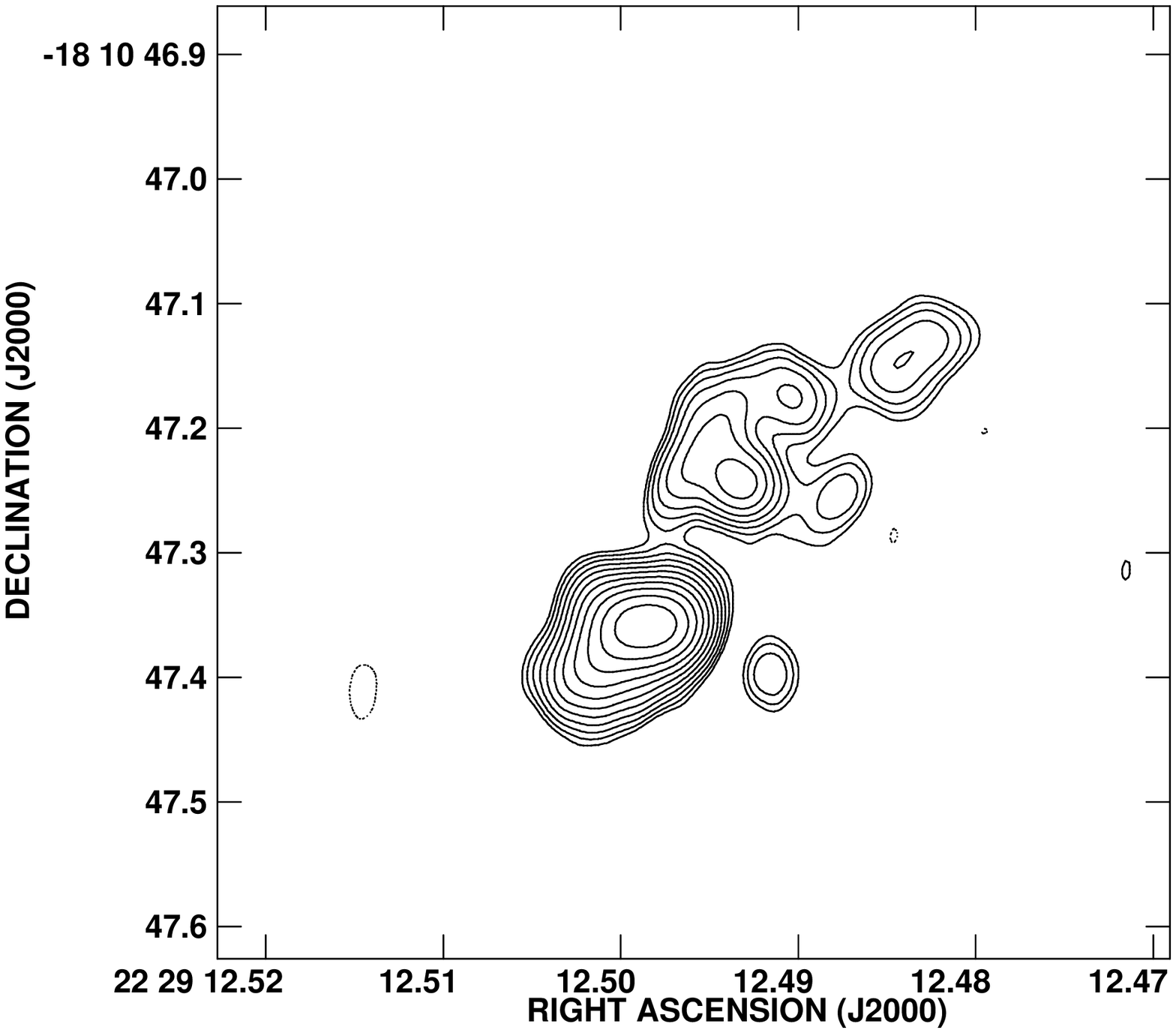}
\caption{The 1.4 GHz continuum emission from \txs.  At left is a
lightly tapered, naturally weighted image with a resolution of 20
$\times$ 12 mas, while at right is a heavily tapered image convolved
with a 45 mas beam.  Contours in the left panel are drawn at $-$0.1,
0.1, 0.14, 0.2, ..., 2.82 mJy beam$^{-1}$ by $\sqrt{2}$ intervals with
negative contours shown dashed.  Contours in the right panel are drawn
at $-$0.2, 0.2, 0.28, 0.4, ..., 11.31 mJy beam$^{-1}$ by $\sqrt{2}$ intervals
with negative contours shown dashed.\label{fig2}}
\end{figure}

%Figure 3 -- Selected Spectra
\begin{figure}
\plotone{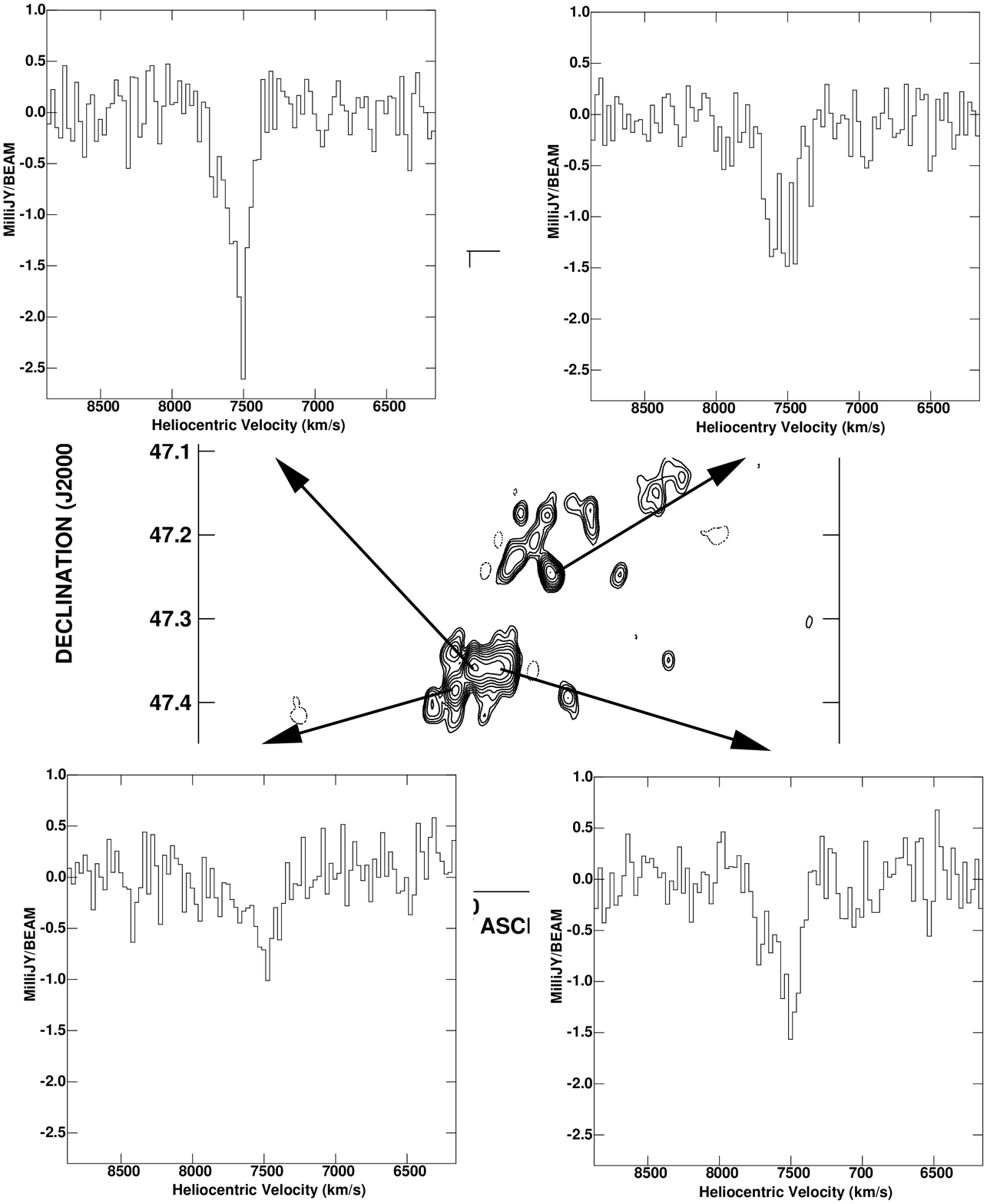}
\caption{Selected spectra of components in \txs\ overlaid on the
radio continuum image from Fig.~2a.  The velocity resolution is
23.98 \kms.  The velocity axis is given in the heliocentric optical
frame ($cz$). \label{fig3}}
\end{figure}

%Figure 4 -- Optical depth
\begin{figure}
\plotone{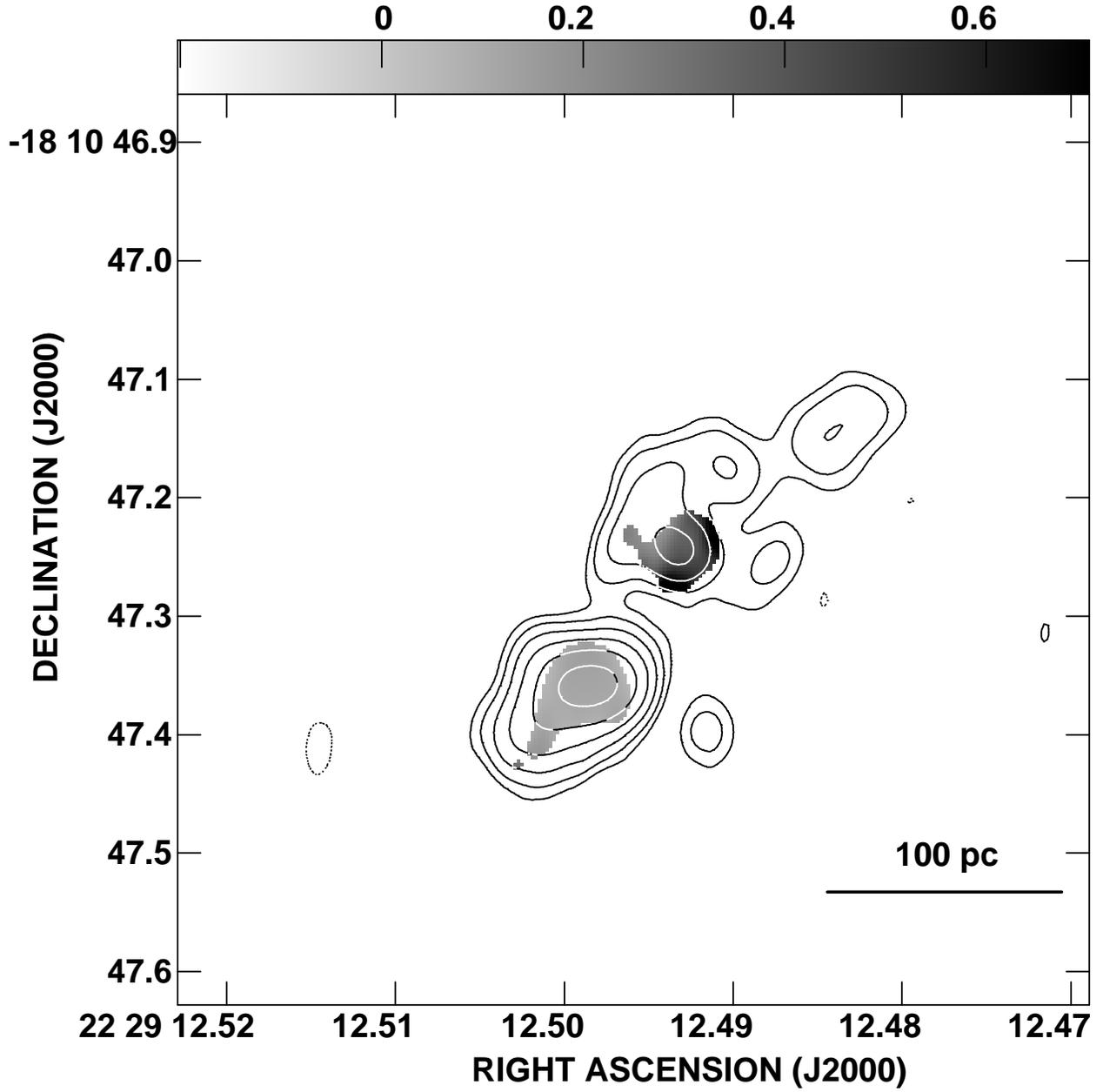}
\caption{A plot of the optical depth ($\tau$) in \txs\ overlaid on the
radio continuum image from Fig.~2b.   A scale bar is drawn in the 
lower right corner.
\label{fig4}}
\end{figure}

%Figure 5 -- Position Velocity
\begin{figure}
\plotone{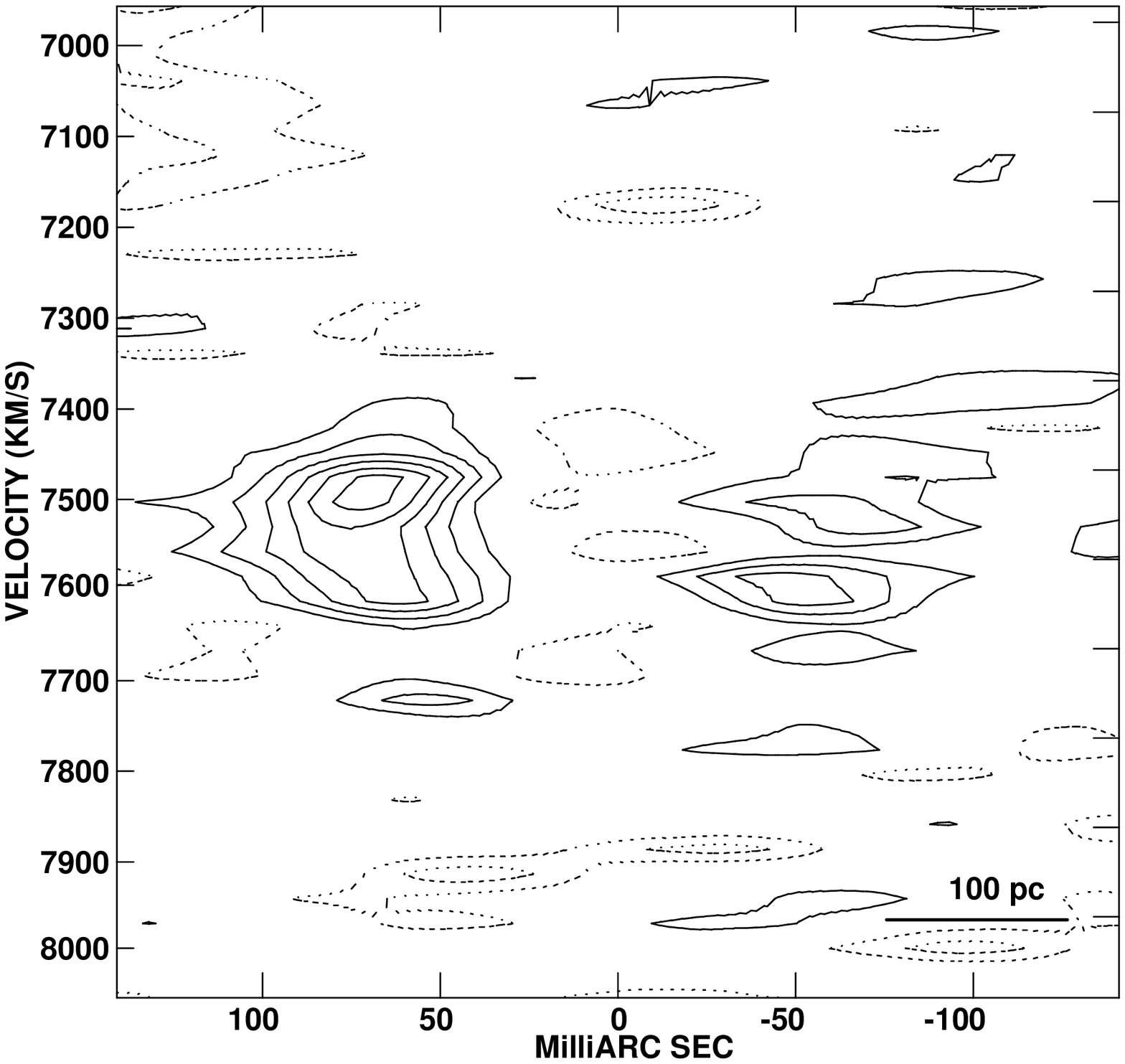}
\caption{A position-velocity plot showing the intensity of the 
absorption in mJy as a function of right ascension, after rotating
the source by $-$55$^\circ$ and averaging in declination over the
source.  Contours are plotted at $-$2.4, $-$2, $-$1.6, $-$1.2, $-$0.8, $-$0.4,
0.4, and 0.8 mJy beam$^{-1}$ with negative contours drawn solid and 
positive contours drawn as dashed lines.
\label{fig5}}
\end{figure}


\clearpage

\begin{references}

\reference{bal04} Ball, G. H., Greenhill, L. J., Moran, J. M., 
Henkel, C., \& Zaw, I. 2004, in preparation

\reference{bes04}Beswick, R. J., Peck, A. B., Taylor, G. B., \& 
Giovannini, G. 2004, MNRAS, in press

\reference{bra96}Braatz, J. A., Wilson, A. S. \& Henkel, C. 1996, ApJS, 
106, 51

\reference{bra97}Braatz, J. A., Wilson, A. S. \& Henkel, C. 1997, ApJS, 
110, 321

\reference{car98}Carilli, C. L., Wrobel, J. M., \& Ulvestad, J. S.\ 1998,
AJ, 115, 928

\reference{car00}Carilli, C.L., \& Taylor, G.B.\ 2000, ApJ, 532, L95

\reference{cla98} Claussen, M.~J., Diamond, P.~J., Braatz, J.~A.,
Wilson, A.~S. and Henkel, C. 1998, ApJL, 500, 129

\reference{con92}Condon, J.~J.\ 1992, ARA\&A, 30, 575 

\reference{con98}Condon, J.J., Cotton, W.D., Greisen, E.W., Yin, Q.F., Perley,
R.A., Taylor, G.B., \& Broderick, J.J.\ 1998, AJ, 115, 1693

\reference{eli95} Elitzur, M. 1995, RMxAC, 1, 85

\reference{fal98}Falcke, H., Wilson, A. S., \& Simpson, C. 1998, ApJ, 502, 199

\reference{fal00}Falcke, H., Wilson, A.S., Henkel, C., Brunthaler, A., 
\& Braatz, J.A.\ 2000, ApJ, 530, L13

\reference{gal96}Gallimore, J. F., Baum, S. A. and O'Dea, C. P. 1996 ApJ, 
464, 198

%\reference{gal99}Gallimore, J. F., Baum, S. A., O'Dea, C. P., Pedlar,
%A. \& Brinks, E. 1999, ApJ, 524, 684

\reference{gio04}Giovannini, G., Taylor, G.B., Cotton, W.D., Feretti, L., 
Lara, L., \& Venturi, T.\ 2004, in preparation

\reference{gre95}Greenhill, L. J., Henkel, C., Becker, R., Wilson, T. L. \& Wouterloot, J. G. A.\ 1995, A\&A, 304, 21

%\reference{gre97}Greenhill, L. J. \& Gwinn, C. R. 1997, Ap\&SS, 248, 261

%\reference{gr297}Greenhill, L. J., Herrnstein, J. R., Moran, J. M., Menten, K. M. \& Velusamy, T. 1997, ApJ, 486, L15

\reference{hec90}Heckman, T. M., Armus, L., \& Miley, G. K.\ 1990, 
ApJS, 74, 833

\reference{her93}Herrnstein, J.~R., et al.\ 1999, Nature, 400, 539

\reference{koe95}Koekemoer, A. M., Henkel, C., Greenhill, L.J., Dey,
A., van Breugel, W., Codella, C., \& Antonucci, R.\ 1995, Nature, 378,
697

%\reference{mar98}Martel, A. R., Baum, S. A., Sparks, W. B., Wyckoff,
%  E., Biretta, J. A., Golombek, D., Macchetto, F. D., McCarthy, P. J.,
%  De Koff, S. \& Miley, G. K. 1998, BAAS, 192, 5204

\reference{mau87} Mauersberger, R., Henkel, C. \& Wilson, T.~L. 1987,
\aap, 173, 352

\reference{mor03} Morganti, R., Osterloo, T. A., Emonts, B. H. C., 
van der Hulst, J. M., \& Tadhunter, C. M.\ 2003, ApJ, 593, L69

\reference{mor04} Morganti, R., et al. 2004, A\&A, in press

\reference{miy95}Miyoshi, M., Moran, J., Herrnstein, J., Greenhill, L.,
Nakai, N., Diamond, P. \& Inoue, M. 1995, Nature, 373, 127

\reference{mye85}Myers, S.T., \& Spangler, S.R. 1985, ApJ, 291, 52

\reference{nef04} Neff, S.G., Ulvestad, J.S., \& Jeno, S. H.\ 2004, ApJ, 
in press

\reference{neu95} Neufeld, D. A. \& Maloney, P. R., \& Conger, S.\ 1994, 
ApJ, 436, 127

%\reference{ode86}O'Dea, C. P. and Owen, F. N. 1986, ApJ, 301, 841

%\reference{ode94}O'Dea, C. P., Baum, S. A., \& Gallimore, J. F. 
1994, ApJ, 436, 669

\reference{pec04} Peck, A. B., Tarchi, A., Henkel, C, et al. 2004, {\it in preparation}

\reference{pec03}Peck, A. B., Henkel, C., Ulvestad, J. S. et al. 2003, ApJ, 590, 149

\reference{ric88}Rice, W., Lonsdale, C.J., Soifer, B.T., Neugebauer, G., 
Kopan, E.L., Lloyd, L.A., de Jong, T., \& Habing, H.J.\ 1988, ApJS, 68, 91

%\reference{sch01}Schmitt, H.~R., 
%Antonucci, R.~R.~J., Ulvestad, J.~S., Kinney, A.~L., Clarke, C.~J., \& 
%Pringle, J.~E.\ 2001, ApJ, 555, 663 

\reference{tar03}Tarchi, A., Henkel, C., Peck, A.B., Nagar, N., Moscadelli, L., \& Menten, K. M.\ 2003, in ``The Neutral ISM in Starburst Galaxies'', 
eds. S. Alto et al., astro-ph/0309446

\reference{tay02}Taylor et al. 2002, ApJ, 574, 88

\reference{wei02}Weiler, K.~W., Panagia, N., Montes, M.~J., \& Sramek, 
R.~A.\ 2002, ARA\&A, 40, 387

\end{references}
\end{document}